\documentstyle[11pt,psfig]{article}

\title{Thermal Restoration of Chiral Symmetry in Supersymmetric
Nambu-Jona-Lasinio Model with Soft SUSY Breaking}
\author{J. Hashida, T. Muta and K. Ohkura\\
Deaprtment of Physics\\
Hiroshima University\\
Higashi-Hiroshima\\
Hiroshima 739-8526}
\date{May 11, 1998}
\begin{document}
\maketitle
\begin{picture}(5,2)
\put(280,280){HUPD-9814}
\end{picture}
\makeatletter
\def\@cite#1#2{$^{\hbox{\scriptsize{#1\if@tempswa , #2\fi})}}$}
\makeatother
\begin{abstract}
     The supersymmetric version of the Nambu-Jona-Lasinio model is investigated
in connection with the chiral symmetry breaking induced by a soft
SUSY breaking term. It is found that the broken chiral symmetry due to the
soft breaking term is restored at suitably high temperature and the symmetry
restoration occurs as first-order phase transitions. The critical
temperature at which the chiral symmetry is restored is determined as a
function of the strength of the soft breaking term and the field coupling
constant. The dynamical fermion mass is calculated at finite temperature. Some
possible applications to the breaking scenario of unified field theories are
discussed.
\end{abstract}

\newpage

     In various theoretical circumstances the Nambu-Jona-Lasinio model \cite{Namb}
plays a role of the low energy effective theory of the more fundamental unified
field theories. Its supersymmetric version is of special interest because the
model is useful to study the low energy consequences of the supersymmetric grand
unified theories. In particular it is important to see whether the chiral symmetry
posessed by the model is broken dynamically to generate fermion masses.
Unfortunately, however, it is well-known \cite{Lov} that the supersymmetry
of the model protects the chiral symmetry from its breaking. If one introduces
a soft SUSY breaking term in the model, one finds that the chiral symmetry in the
model is broken for a sufficiently large soft SUSY breaking term in comparison
with the coupling constant in the interaction term .\cite{Ell} This situation
gives rise to a possibility of studying the phenomenological consequences of the
broken chiral symmetry in the supersymmetric theory.

     In connection with the early history of the universe it is especially
interesting to investigate the finite temperature behavior of the broken chiral
symmetry in the supersymmetric model like the one mentioned above. In the present
communication we shall examine whether the thermal restoration of the chiral
symmetry takes place in the supersymmetric Nambu-Jona-Lasinio model with the soft
SUSY breaking term.

     The Nambu-JonaLasinio model is characterized by the Lagrangian

\begin{equation}
{\cal L}_{NJL}=i\bar{\psi}_{+}\bar{\sigma}^{\mu}\partial_\mu\psi_{+}
	+i\bar{\psi}_{-}\bar{\sigma}^{\mu}\partial_\mu\psi_{-}
	+\frac{G}{N}\psi_{+}\psi_{-}\bar{\psi}_{+}\bar{\psi}_{-},
\end{equation}
where Weyl spinors ${\psi}_{\pm }$ are assumed to consist of $N$
components and we shall work in the leading order of the $1/N$ expansion
throughout the paper.
A natural SUSY extension of the model is given by the Lagrangian \cite{Lov}

\begin{equation}
{\cal L}_{SNJL}=\int d^4\theta\ (\bar{\Phi}_{+}\Phi_{+}+\bar{\Phi}_{-}\Phi_{-}
	+\frac{G}{N}\ \Phi_{+}\Phi_{-}\bar{\Phi}_{+}\bar{\Phi}_{-}),
\end{equation}
where $d^4 \theta=d^2\theta d^2\bar{\theta}$ and chiral superfield
${\Phi}_{\pm }$ is
represented by component scalar ($\phi_{\pm}$) and spinor ($\psi_{\pm}$) fields
and auxiliary field $F_{\pm}$,
\begin{eqnarray}
\Phi_{\pm}(y)=\phi_{\pm}(y)+\sqrt{2}\theta\psi_{\pm}(y)+F_{\pm}(y)\theta^2,
\end{eqnarray}
with $y=x+i\theta\sigma^\mu\bar{\theta}$. The chiral superfields
$\Phi_{+}$ and $\Phi_{-}$ belong to the multiplet {$N$} and {$\bar{N}$} in
SU(N) respectively.

In this SUSY version of the Nambu-Jona-Lasinio model we introduce the following
soft SUSY breaking term \cite{Ell}

\begin{eqnarray}
{\cal L}_{SB}&=&-\int d^4 \theta\ \Delta^2\theta^2 \bar{\theta}^2
(\bar{\Phi}_{+}\Phi_{+}+\bar{\Phi}_{-}\Phi_{-}).
\end{eqnarray}
It is convenient to introduce auxiliary chiral superfield $H$ and $S$ and
use the Lagrangian
equivalent to Eq.(2) with the soft breaking term:

\begin{eqnarray}
{\cal L}
&=&\int d^4\theta [(\bar{\Phi}_{+}\Phi_{+}+\bar{\Phi}_{-}\Phi_{-})
	(1-\Delta^2\theta^2 \bar{\theta}^2)+\frac{N}{G}\bar{H}H]
\nonumber \\
&+&\int d^2\theta\ (\frac{N}{G}HS-S\Phi_{+}\Phi_{-}+h.c.).
\end{eqnarray}
We write the relevant Lagrangian in terms of component fields to
calculate the effective potential in the leading order of $1/N$ expansion:
\begin{eqnarray}
{\cal L}&=&-\bar{\phi}_{+}(\Box +|\phi_S|^2+\Delta^2)\phi_{+}
	-\bar{\phi}_{-}(\Box +|\phi_S|^2+\Delta^2)\phi_{-} \nonumber\\
&&	\hspace{-0.8cm}+i\bar{\psi}_{+}\bar{\sigma}^\mu
	\partial_\mu\psi_{+}+i\bar{\psi}_{-}\bar{\sigma}^\mu
	\partial_\mu\psi_{-} 
	+\psi_{+}\psi_{-}\phi_S +\bar{\psi}_{+}\bar{\psi}_{-}\bar{\phi}_S
	-\frac{N}{G}|\phi_S|^2.
\end{eqnarray}
where we have used equations of motion for the auxiliary fields 
$F_{\pm}\ $and$\ F_H$.
The field $\phi_S$ is related to the ordinary scalar ($\sigma$) and pseudoscalar
($\pi$) field such that
\[
\phi_S=\sigma + i \pi.
\]
The effective potential for this Lagrangian is easily written down in the leading
order of the $1/N$ expansion and reads

\begin{equation}
V_0(|\phi_S|^2)=\frac{|\phi_S|^2}{G}+2i\int \frac{d^4k}{(2\pi)^4}(\ln 
	\frac{-k^2+|\phi_S|^2}{-k^2}
	-\ln\frac{-k^2+|\phi_S|^2+\Delta^2}{-k^2+\Delta^2}). 
\end{equation}
After Wick rotation we perform the $k_0$ integration and obtain
\begin{eqnarray}
V_0(|\phi_S|^2)=\frac{|\phi_S|^2}{G}
&&\hspace{-0.6cm}+2\int\frac{d^3k}{(2\pi)^3}
	[\ \sqrt{k^2+|\phi_S|^2+\Delta^2}-\sqrt{k^2+|\phi_S|^2}\nonumber\\
&&	-(\sqrt{k^2+\Delta^2}-\sqrt{k^2})\ ].
\end{eqnarray}
Note here that the effective potential is normalized so that $V_0(0)=0$.
By analyzing the above effective potential we recognize that the chiral symmetry
is broken for sufficiently large $\Delta$. This situation is understood more
clearly by observing the gap equation which is derived from the effective
potential (8). The gap equation $\partial V_0/\partial |\phi_S|^2=0$ is found to be

\begin{equation}
	\frac{1}{G}=\int\frac{d^3k}{(2\pi)^3}
	(\frac{1}{\sqrt{k^2+|\phi_S|^2}}-\frac{1}{\sqrt{k^2+|\phi_S|^2+\Delta^2}}).
\end{equation}
Introducing the three dimensional cut-off $\Lambda$ in the momentum integration
we perform the integration in Eq. (9) and obtain

\begin{eqnarray}
\frac{1}{2\alpha}&=&\sqrt{1+x}-\sqrt{1+x+\delta}
	+(x+\delta)\ln\frac{1+\sqrt{1+x+\delta}}{\sqrt{x+\delta}}\nonumber\\
&&	-x \ln\frac{1+\sqrt{1+x}}{\sqrt{x}},
\end{eqnarray}
where $\alpha$, $\delta$ and $x$ are defined by
\begin{eqnarray}
&&	\alpha=\frac{G\Lambda^2}{8\pi^2},\
	\delta=\frac{\Delta^2}{\Lambda^2},\ 
	x=\frac{|\phi_S|^2}{\Lambda^2}.
\end{eqnarray}
It is not difficult to show that Eq. (10) allows a nontrivial solution for
$x$ with the suitable choice of parameters $\alpha$ and $\delta$.
Thus the dynamical fermion mass is generated and the chiral symmetry is broken
dynamically.
Noting that the chiral symmetry breaking in the present case is of the second
order phase transition we let the gap parameter $x$ vanish so that we obtain
the critical $\delta$ at which the chiral symmetry is broken. The critical
$\delta$ is given by solving the following relation as a function of the coupling
constant $\alpha$:

\begin{equation}
\frac{1}{2\alpha}=1-\sqrt{1+\delta}+\delta\ln
	\frac{1+\sqrt{1+\delta}}{\sqrt{\delta}}
\end{equation}

     Now we introduce the temperature. We employ the Matsubara formalism
\cite{Mat} to introduce temperature effects in our calculation of the
effective potential.
We simply replace the integral in energy variable by the summation where we
apply the periodic boundary condition for boson fields and the anti-periodic
boundary condition for fermion fields respectively. The energy variable $k_0$
in the summation is replaced in the following way:
\begin{equation}
\int\frac{dk_0}{2\pi}f(k_0)\rightarrow\frac{1}{\beta}
\sum_{n=-\infty}^{\infty}f(\omega_n),
\end{equation}
where $\beta$ is the inverse of the temperature and $\omega_n$ 
is given by 
\[
\omega_n = \left\{
\begin{array}{ll}
		\omega_n^B\equiv\frac{2n}{\beta}\pi&(\textrm{boson}),\\
\\
		\omega_n^F\equiv\frac{2n+1}{\beta}\pi&(\textrm{fermion}).
\end{array}
\right.
\]
The calculation is essentially the same as the one described in the previous
paper.\cite{Ina}
We first calculate the effective potential for the supersymmetric
Nambu-Jona-Lasinio model without the soft SUSY breaking term and obtain
\begin{equation}
V(|\phi_S|^2)=\frac{|\phi_S|^2}{G}-\frac{2}{\beta}
	\sum_{n}\int \frac{d^3k}{(2\pi)^3}\ 
	(\ln \frac{{\omega_n^F}^2+k^2+|\phi_S|^2}{{\omega_n^F}^2+k^2}
	-\ln\frac{{\omega_n^B}^2+k^2+|\phi_S|^2}{{\omega_n^B}^2+k^2}).
\end{equation}
The summation on $n$ in Eq. (14) is performed easily to give
\begin{equation}
V(|\phi_S|^2)=\frac{|\phi_S|^2}{G}
	-\frac{4}{\beta}\int\frac{d^3k}{(2\pi)^3}
	\ln\ [ \coth\frac{\beta\sqrt{k^2+|\phi_S|^2}}{2}
	\tanh\frac{\beta\sqrt{k^2}}{2}\ ].
\end{equation}
The result reflects the well-known fact that the supersymmetry is spoiled
at finite temperature according to the different distribution functions for
bosons and fermions.\cite{Ojim} It is, however, interesting to note that the
supersymmetry is preserved in the low temperature region where the Boltzmann
distribution is a good approximation. In fact, if we keep only the dominant terms
for small $T$ in Eq.(15), we find that the second term in Eq.(15) disappears
leaving the tree expression of the effective potential. We realize that the
supersymmetry is a good symmetry only in the low temperature region.

     We then take into account the soft SUSY breaking term and calculate the
effective potential to obtain

\begin{eqnarray}
V(|\phi_S|^2)=\frac{|\phi_S|^2}{G}-&&\hspace{-0.6cm}\frac{2}{\beta}
	\sum_{n}\int \frac{d^3k}{(2\pi)^3}\ 
	(\ln \frac{{\omega_n^F}^2+k^2+|\phi_S|^2}{{\omega_n^F}^2+k^2}
	\nonumber\\
	&&-\ln \frac{{\omega_n^B}^2+k^2+|\phi_S|^2+\Delta^2}
	{{\omega_n^B}^2+k^2+\Delta^2}).
\end{eqnarray}
The summation on $n$ in the above equation can be easily performed
as before and results in
\begin{equation}
V(|\phi_S|^2)=V_0(|\phi_S|^2)
	-\frac{4}{\beta}\int\frac{d^3k}{(2\pi)^3}
	\ln(\frac{1+e^{-\beta\sqrt{k^2+|\phi_S|^2}}}{1+e^{-\beta\sqrt{k^2}}}
	\frac{
	1-e^{-\beta\sqrt{k^2+\Delta^2}}}
	{1-e^{-\beta\sqrt{k^2+|\phi_S|^2+\Delta^2}}}),
\end{equation}
where $V_0$ is the effective potential for vanishing temperature already
given in Eq.(8).

     For almost whole range of the set of parameters $\alpha$ and $\delta$
we estimated the effective potential (17) by numerical integration. We show
a typical example of its behavior in Fig. 1 for several values of $\beta$.
\\
\\
\hspace*{6cm}- Fig.1 -
\\
\\
Apparently we observe that the broken chiral symmetry at $T=0$ is restored
at critical temperature $\beta\Lambda=13$ and
the chiral symmetry restoration occurs as a first-order phase transition for
which the mass gap is expected at the critical temperature. In order to see the
situation more clearly we write down the gap equation
\begin{eqnarray}
\frac{1}{G}&=&\int\frac{d^3k}{(2\pi)^3}
	(\frac{1}{\sqrt{k^2+|\phi_S|^2}}-\frac{1}{\sqrt{k^2+|\phi_S|^2+\Delta^2}})
\nonumber\\
&&-2\int\frac{d^3k}{(2\pi)^3}[\ \frac{1}{\sqrt{|\phi_S|^2+k^2}}
	\frac{1}{e^{\beta\sqrt{|\phi_S|^2+k^2}}+1}\nonumber\\
&&	-\frac{1}{\sqrt{|\phi_S|^2+k^2+\Delta^2}}
	\frac{1}{e^{\beta\sqrt{|\phi_S|^2+k^2+\Delta^2}}-1}\ ].
\end{eqnarray}

     By the numerical anlysis of Eq.(18) we recognize that two nontrivial
solutions for $x$ develop as temperature goes through the critical value
signaling the possibility of
the first-order phase transition. The critical temperature $T_C$ and the
critical value of $x$ are determined by solving the gap equation (18) and
$V=0$ simultaneously.
The critical temperature $T_C$ is obtained at the same time as in the above
analysis and is a function
of the soft breaking parameter $\delta=\Delta^2/\Lambda^2$ for a fixed
$\alpha=G\Lambda^2/8\pi^2$.

     In the broken phase $x$ is non-vanishing and hence there develops the
non-vanishing vacuum expectation value of the scalar field $\sigma$ which
is responsible for generating fermion masses. In Fig. 2 we present the 
dynamical fermion mass $m/\Lambda=\sqrt{x}$ as a function of 
$\beta\Lambda$. We clearly observe the mass gap characteristic of the 
first-order phase transition.
\\
\\
\hspace*{6cm}- Fig.2-
\\
\\
It should be pointed out here that the integration in Eq. (18) is performed
rigorously in the case of the three dimensional space-time. We leave
the argument in the three dimensional case in the forthcoming publication.

The supersymmetric four-fermion theory is considered to be a low-energy
manifestation of the more fundamental supersymmetric unified theories. The
typical example is the supersymmetric GUT which breaks down to the electroweak
theory. In the early universe scenario this stage of the breaking of the
supersymmetric GUT corresponds to the beginning of the inflation era where
the first-order phase transition is required. In this connection our finding
of the chiral symmetry restoration as the first-order phase transition is
of great interest.

     The authors would like to thank Dr. T. Inagaki and Dr. S. Mukaigawa for
valuable comments and discussions. One of the authors (K. O) would like
to express his sincere gratitude to Dr. T. Onogi for the guidance in the study of
supersymmetric theories. The other of the authors (T. M.) is supported by
the Monbusho Grant-in-Aid for Scientific Research (C) No.08640377 and for
Scientific Research on Priority Areas No.09246105.

Figure Captions
\\
\\
Fig 1:\ Effective potential at finite temperature for
        $G\Lambda^2=7 \times 10^{3}$ and\\
\              $\Delta/\Lambda=0.1$
\\
\\
Fig 2:\ Dynamical mass of fermion as a function of 
           $\beta\Lambda(\beta_C\Lambda=\Lambda/T_C=13)$
\newpage 

\begin{figure}[hbpt]
  \begin{center}
    \psfig{file=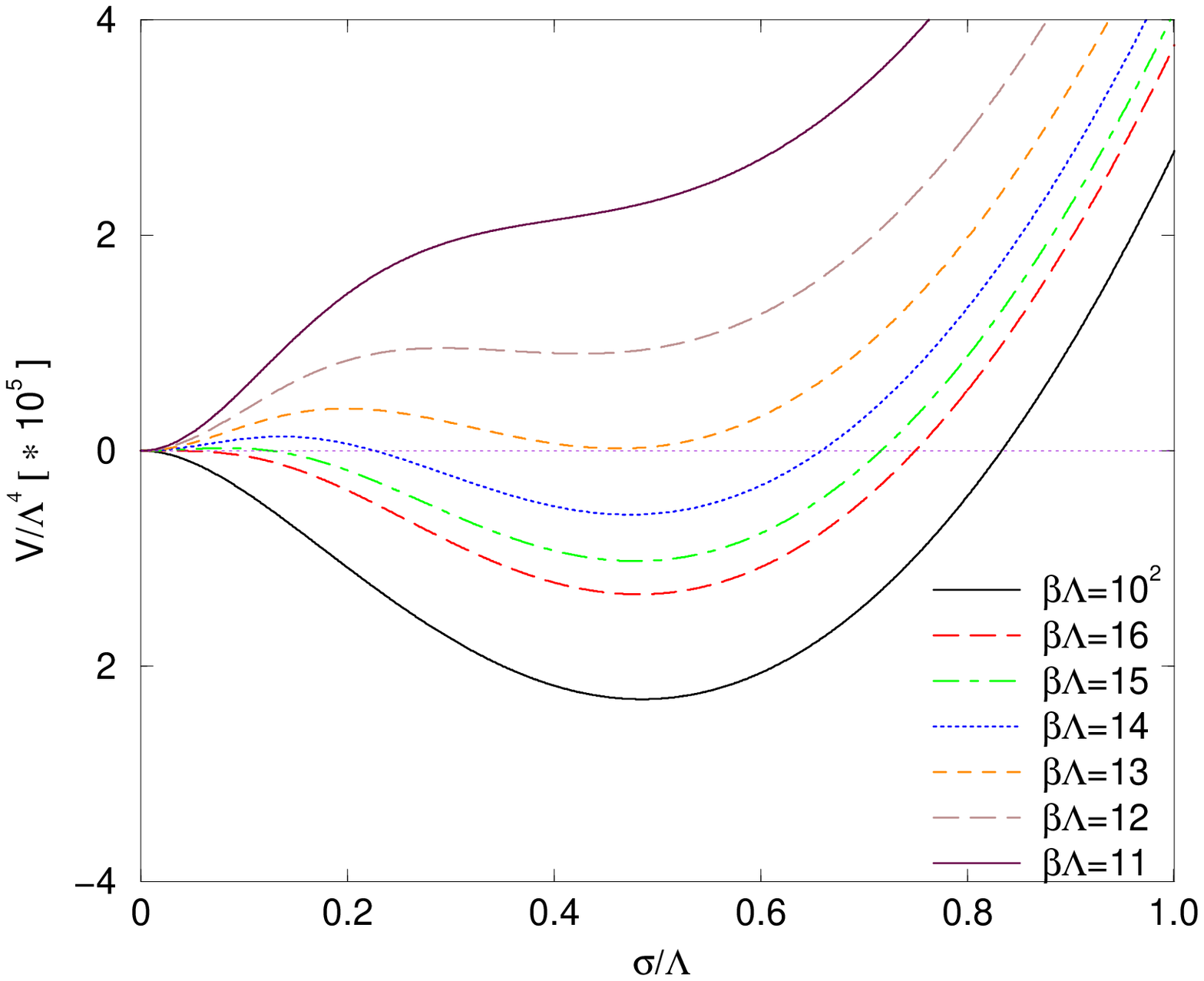,height=15cm,width=15cm}
    \caption{} 
  \end{center}
\end{figure}

\newpage 

\begin{figure}[hbpt]
  \begin{center}
    \psfig{file=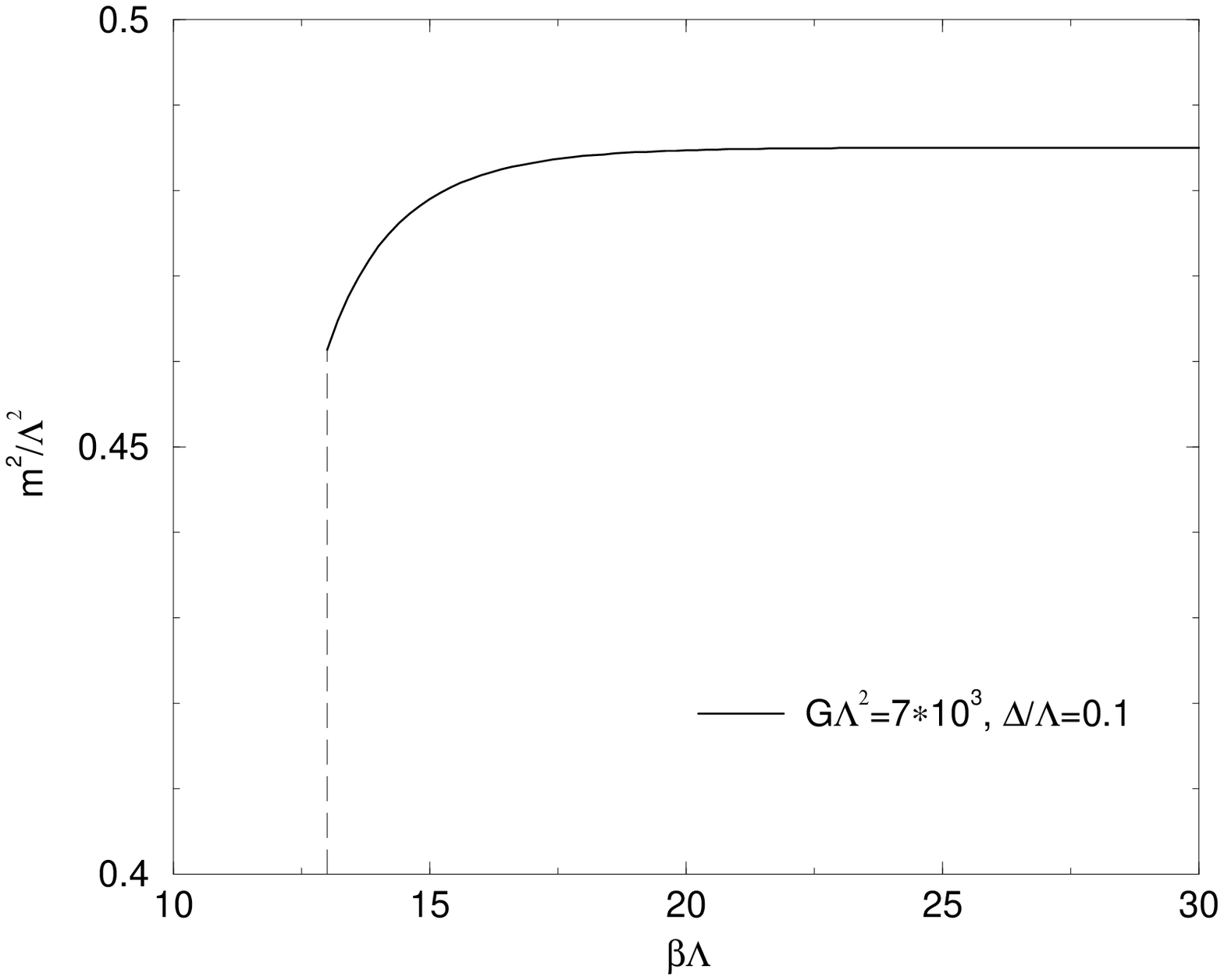,height=15cm,width=15cm}
    \caption{} 
  \end{center}
\end{figure}

\end{document}